\documentstyle[twoside,fleqn,epsf,espcrc2]{article}

\newcommand{\AmS}{{\protect\the\textfont2
  A\kern-.1667em\lower.5ex\hbox{M}\kern-.125emS}}

\title{Scaling Study of the Leptonic Decay Constants of Heavy-Light Mesons:
       A Consumers Report on Improvement Factors }

\author{S. G\"usken\address{Physics Department, University of Wuppertal,
        D-42097 Wuppertal, Germany}, in collaboration with \\
        C. Alexandrou\address{Department of Natural Sciences,
        University of Cyprus, Nicosia, Cyprus},
        F. Jegerlehner\address{PSI, CH-5232 Villigen, Switzerland},
        K. Schilling$^a$, G. Siegert$^a$ and
        R. Sommer\address{DESY, Theory Division, D-22603 Hamburg, Germany}}

\begin{document}

\begin{abstract}
A high statistics calculation, performed at $\beta =5.74,\;6.00$ and $6.26$,
 enables us to study the variation of the leptonic decay constants $f_P$ of
heavy pseudoscalar mesons with the lattice spacing $a$.
We observe only a  weak $a$ dependence when
the standard $\sqrt{2\kappa}$ normalization is used for the quark fields,
whereas application of  the Kronfeld-Mackenzie normalization induces
a stronger variation with $a$.
Increasing the meson mass from $1.1GeV$ to $2.3GeV$
this situation becomes even more pronounced.
\end{abstract}

\maketitle

\section{INTRODUCTION}

The prediction of the leptonic decay constants of the $D$ and the
$B$ meson within the framework of lattice QCD is a challenging but
also very delicate problem, since
in the region of heavy mesons, the inverse of the respective
  masses comes close to
currently reachable lattice resolutions. Therefore
large discretization effects may contaminate the results .

The question how to suppress these unphysical contributions has been
tackled from various sides\cite{Seiswohl,BLS,LKroMac}.
Using meanfield arguments, Kronfeld and Mackenzie suggested
 that the replacement of the  standard $\sqrt{2\kappa}$ normalization
of Wilson quark fields by $\sqrt{1-3 \kappa/4\kappa_c}$ should signifcantly
reduce the effects of finite lattice spacing.

The present study takes a rather empirical approach to the issue of finite
$a$ effects in the region of heavy mesons. We vary the lattice spacing
in the currently accessible range and analyse the corresponding variation
of $f_P$. Clearly a strong dependence of $f_P$ on $a$ would indicate the
presence of large discretisation errors, whereas a weak dependence
would be associated with smaller contaminations.
In this sense our approach is perfectly suited to judge on the efficiency
of different quark field normalizations within a given $a$ region.

We will finally perform an extrapolation of $f_P$ to the continuum, assuming
that its functional dependence on $a$ is linear in the leading part, as
suggested by the data.

\section{PREPARATION}

\begin{table*}[hbt]
\setlength{\tabcolsep}{1.5pc}
\newlength{\digitwidth} \settowidth{\digitwidth}{\rm 0}
\catcode`?=\active \def?{\kern\digitwidth}
\caption{Lattice parameters}
\label{setup}
\begin{tabular*}{\textwidth}{llllll}
\hline
                  \multicolumn{2}{l}{$\beta=5.74,\; a^{-1}_{\sigma}=1.118(9)$}
                 & \multicolumn{2}{l}{$\beta=6.00,\;
a^{-1}_{\sigma}=1.876(19)$}
                 & \multicolumn{2}{l}{$\beta=6.26,\;
a^{-1}_{\sigma}=2.775(18)$}\\
\cline{1-2} \cline{3-4} \cline{5-6}

$N_{S},\;N_{T}$ & $no.confs.$ &$N_{S},\;N_{T}$ &$no.confs.$ & $N_{S},\;N_{T}$
& $no.confs.$ \\
\hline
4,24   & 404 & 6,36 & 227 &        &       \\
6,24   & 131 &      &     & 12,48  & 103   \\
8,24   & 175 & 12,36& 204 & 18,48  & 76    \\
10,24  & 213 & 18,36& 9   &        &       \\
\hline
\end{tabular*}
\end{table*}
In order to visualize unambiguously the $a$ dependence of $f_P$, we have to
take care that the finite $a$ effects are not hidden in the statistical
noise or distorted by incomplete groundstate projection of the
meson propagator and effects due to the finite size of the lattice.
Therefore we have done our calculation with high statistics, keeping
the errors of the raw data below $5\%$. We have varied the lattice size
from about 0.7~fm to 2~fm and have smeared the quark fields with the
well established  \cite{fb3}
Gauss like Wuppertal wavefunction ($n=100,\; \alpha=4$).
In table \ref{setup} we display the lattice parameters together
with the lattice spacing, taken from the stringtension $\sigma$ \cite{Balisch}.
The influence of the finite lattice extension on $f_P$ was checked by
comparing the results at different lattice sizes and fixed lattice
constant.
We find that finite size effects are small once
the lattice extension becomes as large as $1.4 fm$.

\section{RESULTS}

\subsection{Finite $a$ effects}

\begin{figure}[htb]
\epsfxsize=7.0cm
\epsfbox{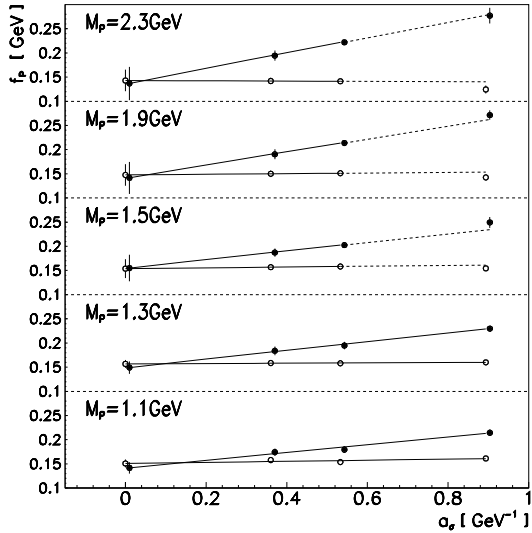}
\caption{$f_P$ as a function of $a$. Points connected only by dashed lines
were not used in the extrapolation.}
\label{finita}
\end{figure}

In fig.\ref{finita} we show the leptonic decay constant $f_P$ as a
function of $a$ both in the $\sqrt{2\kappa}$ normalization (open symbols) and
in the Kronfeld-Mackenzie  normalization (closed symbols).
The light quark mass has been extrapolated to the chiral limit and
we have interpolated between the results at
 adjacent heavy quark masses (c.f. table \ref{extrap})
 in order to keep the meson mass (in GeV) fixed when the $a$ dependence
of $f_P(M_P,a)$ is investigated.
 Due to their small statistical errors, we have
 used stringtension measurements \cite{Balisch}
to relate our data to a physical scale.
 The renormalization factor $Z_A$ was taken from perturbation theory
 \cite{Smit}\footnote{ $Z_A = 1-0.1333g^2$ for standard normalization
and $Z_A = 1-0.0248g^2$ in the case of KroMac normalization.}, with
 an effective coupling $\tilde{g}^2=3 g_0^2/<tr P_{\mu \nu}>$,
recommended in ref.\cite{LKroMac} ($P_{\mu \nu}\equiv$ 1$\times$1
 Wilson loop).

It goes without saying that the $a$ dependence of $f_P$ must be different
in the two normalizations. Very surprisingly, however, fig.\ref{finita}
shows clearly that - in contrast to the $\sqrt{2\kappa}$ normalized results -
the variation  with $a$ becomes stronger and stronger with increasing
meson mass when the KroMac normalization is used. This means that -- at least
in
the displayed $a$ and $M_P$ range -- the KroMac normalization does a
bad job: Instead of suppressing finite $a$ effects it enhances them.

In order to connect our results to the (physical) continuum, we
followed the behavior suggested by the data in both normalizations
and extrapolated linearly\footnote{Since the $a$ dependence
 cannot be exactly linear for both normalizations at the same
time, we have excluded those points from the fit where
 $\sqrt{1 - 3\frac{\kappa}{4\kappa_c}}/\sqrt{2\kappa} > 1.6$.}to $a=0$.
As can be seen from fig.\ref{finita},we obtain nice
 agreement of the results, although the
KroMac normalization has induced considerably larger errors.
\begin{table*}[hbt]
\setlength{\tabcolsep}{0.57pc}
\catcode`?=\active \def?{\kern\digitwidth}
\caption{Decay constant and meson mass in lattice units. The light
quark has been extrapolated to $\kappa_c$. }
\label{extrap}
\begin{tabular*}{\textwidth}{lllllllll}
\hline
\multicolumn{3}{l}{$\beta=5.74$} &
\multicolumn{3}{l}{$\beta=6.00$} &
\multicolumn{3}{l}{$\beta=6.26$} \\
\hline
$\kappa_h$ & $f_P/Z_A$ & $M_P$ &
$\kappa_h$ & $f_P/Z_A$ & $M_P$ &
$\kappa_h$ & $f_P/Z_A$ & $M_P$ \\
\hline
   0.06 & 0.1197(102) & 2.502(13)
 & 0.10 & 0.0873(17)  & 1.498(10)
 & 0.09 & 0.0437(32)  & 1.579(19) \\
   0.09 & 0.1629(52)  & 1.871(13)
 & 0.115& 0.0983(18)  & 1.197(7)
 & 0.10 & 0.0486(31)  & 1.375(14) \\
   0.125& 0.1890(33)  & 1.205(7)
 & 0.125& 0.1038(18)  & 0.995(5)
 & 0.120& 0.0609(33)  & 0.965(10) \\
   0.140& 0.1907(38)  & 0.904(6)
 & 0.135& 0.1085(19)  & 0.780(7)
 & 0.135& 0.0689(29)  & 0.636(6) \\
   0.150& 0.1829(53)  & 0.684(5)
 & 0.145& 0.1032(31)  & 0.551(2)
 & 0.145& 0.0711(24)  & 0.382(4) \\
        &             &
 &      &             &
 & 0.1492& 0.0580(50)  & 0.245(4) \\
\hline
\end{tabular*}
\end{table*}

\subsection{Heavy mass extrapolation}

The most 'natural' scale for $f_p$ is $f_{\pi}$,
since the  uncertainty originating from the renormalization constant $Z_A$
 cancels out in this case.
Lattice measurements of $f_{\pi}$ are generally affected with
large statistical errors and therefore
we have decided  to convert our results to this scale only after
having performed the
$a \rightarrow 0$ extrapolation of $f_P$.
To achieve this we have decoupled the extrapolations
according to
$   \frac{f_P}{f_{\pi}}(a \rightarrow 0) =
      \frac{{f_P}/{\sqrt{\sigma}}(a \rightarrow 0)}
           {{f_{\pi}}/{\sqrt{\sigma}}(a \rightarrow 0)}\quad .
$
Although the $O(\tilde{g}^4)$--uncertainty in $Z_A$ does not cancel out
exactly if one first extrapolates and then takes the ratio, its effect
should be roughly the same in numerator and denominator.
To obtain the  denominator of this ratio we used
both  our own data and  the results quoted
in refs.~\cite{GF11,Ape,BLS,UKQCD_light}.
Since the $a$ dependence of ${f_{\pi}}/{\sqrt{\sigma}}$ is
weak, a linear extrapolation to $a=0$ is well justified and leads to
$\frac{f_{\pi}}{\sqrt{\sigma}}(a=0) = 0.269(12)$.

\begin{figure}[htb]
\epsfxsize=7.0cm
\epsfbox{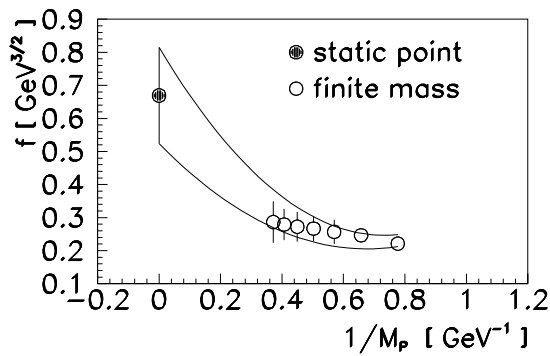}
\caption{ $ \hat{f}$ as a function of $1/M_P$.}
\label{fhat}
\end{figure}

In figure~\ref{fhat} we display our final
results\footnote{The extrapolation has been performed on the data
 in the $\sqrt{2\kappa}$ norm
 since it involves smaller statistical errors than
using the KroMac norm.}  at $a=0$ in the
form\footnote{$\hat{f} = f_P \sqrt{M_P}\times (\frac{\alpha_s(M_P)}
{\alpha_s(M_B)})^{6/33}$ .}
 $\hat{f}_P(1/M_P)$, together with our static value from  ref.~\cite{fb3}.
The new data appears to depend only weakly on $M_P$. Because of the
various extra- and interpolations however, the data points carry
error bars of order $25 \%$ and
therefore do not exclude a stronger variation in $M_P$.
Given this  situation  we draw an error band that links the conventional
results with the static point. The $M_P$ dependence of the error band was
chosen according to the ansatz
$
   \hat{f}_P = c_0 +\frac{c_1}{M_P} + \frac{c_2}{M_P^2}\quad .
$
At the location of the B and D meson the error band corresponds to the
bounds
$
   155 MeV \le f_B \le 242 MeV \;,\; 150 MeV \le f_D \le
   200 MeV \;.
$
It is evident from figure~\ref{fhat} that these bounds
are strongly affected by the size and uncertainty of $f_{\mbox{stat}}$.
More work is necessary to obtain an accurate prediction for $f_B$.

\end{document}